\documentclass[reprint,amsmath,amssymb,aps,superscriptaddress]{revtex4-2}
\usepackage{graphicx}
\usepackage{dcolumn}
\usepackage{bm}
\usepackage{xcolor}
\usepackage{hyperref}
\usepackage{enumitem}
\usepackage[none]{hyphenat}
\DeclareMathAlphabet\mathbfcal{OMS}{cmsy}{b}{n}

\begin{document}
	\title{Filamentation of Femtosecond Vector Beams}
	 
	\author{Amirreza Sadeghpour}
	\affiliation{
		Department of Physics, Institute for Advanced Studies in Basic Sciences (IASBS), Zanjan 45137‑66731, Iran
	}
	\author{Daryoush Abdollahpour}
	\email{dabdollahpour@iasbs.ac.ir}
	\affiliation{
		Department of Physics, Institute for Advanced Studies in Basic Sciences (IASBS), Zanjan 45137‑66731, Iran
	}
	
	\date{\today}

\begin{abstract}
\noindent We numerically study the filamentation of femtosecond vector beams with spatially varying polarization profiles in air. The vector beams include azimuthal, radial, and spiral cylindrical vector beams (CVBs), as well as low-order Full Poincar\'e beams (FPBs) with star and lemon polarization topologies. Comparing the filamentation of CVBs to that of a circularly polarized Laguerre-Gaussian (LG) beam, we found that CVBs undergo filamentation more easily due to their more effective nonlinear focusing, in contrast to the LG beam. In the case of low-order FPBs, we examined the role of the constituent orthogonally polarized spatial modes in their filamentation process. Our findings revealed that the Gaussian mode within these beams primarily contributes to their filamentation. Additionally, when compared to a linearly polarized Gaussian beam, star and lemon FPBs displayed greater resistance to filamentation. Furthermore, we investigated the evolution of polarization profiles of the beams during filamentation. We observed that the polarization profiles of CVBs remained largely unchanged. In contrast, the polarization profiles of the FPBs underwent significant non-uniform changes due to differences in accumulated nonlinear phase and Gouy phase among the constituent modes.     
\end{abstract}
	
	\maketitle
	
	\section{\label{sec1} Introduction}
	Filamentation of femtosecond pulses is a captivating phenomenon that has emerged as a fascinating area of research in the field of ultrafast optics. This unique phenomenon involves the propagation of intense ultrashort laser pulses through transparent materials, where the pulse undergoes self-focusing and self-channeling, leading to the formation of long, stable, and self-guided filaments of light \cite{BraunOL95}. These filaments can stretch over significant distances, defying conventional diffraction limits and maintaining their spatial coherence \cite{NibbOL96}. Since the first experimental observations of self-channeling of high-peak-power femtosecond pulses in air \cite{BraunOL95, NibbOL96}, filamentation has been under extensive investigation to explore the underlying physics or potential applications of this intriguing phenomenon \cite{CouaironPhysRep07}. Among several models for the description of the phenomenon, filamentation is often successfully described as a dynamic equilibrium involving several linear and nonlinear effects including Kerr self-focusing, plasma defocusing, diffraction, dispersion, and losses due to multiphoton absorption and absorption by plasma \cite{MlejOL98, MlejPRL99, BergePRL01, CouaironPRL02}. Filamentation has several important features, including intensity clamping (at $\sim$$10^{13}\ \text{W/cm}^{2}$) over an almost constant diameter persisting over remarkably long distances beyond several Rayleigh ranges, generation of underdense plasma channels, supercontinuum generation, and conical emission \cite{CouaironPhysRep07, TzortzakisPRL01}. 	

	Owing to these features, filamentation has been utilized in numerous applications such as while-light LIDAR \cite{KaspSci03}, single-cycle pulse generation in near- \cite{CouaironOL05}, and mid-IR \cite{MitroOptica16}, high-harmonic generation \cite{SteinNJP11}, micro-/nano-structuring in the bulk of transparent solids \cite{GattassNP08, BharPRL06, PapazAPA14}, cancer treatment \cite{MeesatPNAS12}, generation of plasma photonic devices \cite{SuntsovAPL09}, THz generation \cite{DAmicoPRL07, ZhangPRL16, KoulNC20}, and laser-guided lightning \cite{HouardNP23}.

	In spite of the extensive investigation of filamentation, simplified scalar models are often utilized for the description of filamentation \cite{CouaironPhysRep07}, and the effect of polarization and its dynamics during nonlinear propagation have been mostly unexplored. Present reports on these aspects of filamentation are limited to optical beams with uniform linear, circular, or elliptical polarization \cite{KolesikPRE01, RostamiOE15}. On the other hand, the generation of novel forms of structured light, including vector beams with space-varying polarization profiles, has led to intriguing possibilities for diverse applications and fundamental science \cite{RubinJoP16, HeLSA22}. Cylindrical vector beams (CVBs) \cite{ZhanAOP09}, and Full Poincar\'e beams (FPBs) \cite{BeckleyOE10} are two interesting categories of vector beams. The polarization morphology of the CVBs, which possesses a cylindrical symmetry around the beam axis, is in the form of radial and azimuthal distributions or their superpositions. More interestingly, the polarization distribution in Full Poincar\'e beams spans the entire surface of the Poincaré sphere. Both CVBs and FPBs can be expressed as a superposition of the Laguerre-Gaussian (LG) spatial eigenmodes with orthogonal circular polarizations \cite{BouchardPRL16, GibsonPRA18}. CVBs and FPBs have been used for various applications including optical manipulation \cite{ZhanAOP09, BheSR18}, optical microscopy \cite{ChenOL13}, quantum entanglement of complex photon polarization patterns \cite{FicklerPRA14}, single-shot polarimetry imaging \cite{SivanOL16}, formation of polarization speckle \cite{SallaOE17} and laser-induced mass transport \cite{GarciaOE20}. Moreover, nonlinear propagation of several continuous wave (CW) CVBs and FPBs, in a saturable Kerr medium (e.g., Rb vapor) has been studied in the past few years to investigate the effect of polarization shaping on optical collapse \cite{BouchardPRL16}, polarization rotation of the FPBs \cite{GibsonPRA18}, and suppression of the formation of the optical rouge waves with Poincar\'e beams \cite{BlackPRL22}. Furthermore, electronic photocurrent shaping through polarization shaping of the incident vector beams for THz magnetic field generation \cite{SederbergPRX20, JanaNP21}, and the generation of reconfigurable THz metasurfaces \cite{JanaNanophotonics22} has been recently demonstrated. Most importantly, investigation of the effect and evolution of the polarization distribution during the filamentation of vector beams are of critical importance for several novel applications of the vector beams involving laser-plasma interactions, such as THz generation, harmonic generation, generation of sculptured electronic photocurrent circuits, and also for polarization-sensitive filament-induced material structuring \cite{ManouOL20, RekAOM16, ChengMaterials22, SchilleNanomaterials22}. 
	
	In this study, we perform numerical investigations on the filamentation of femtosecond vector beams, including azimuthal, radial, and spiral CVBs, as well as FPBs with ``star" and ``lemon" polarization topologies. We aim to explore how these polarization patterns affect the nonlinear propagation of the beams. We make comparisons between CVBs and a circularly polarized LG beam, as well as between FPBs and a linearly polarized Gaussian beam. Our findings reveal that, even in an isotropic medium like air, contributions from orthogonal polarization components significantly influence the filamentation of vector beams. Specifically, CVBs are more susceptible to filamentation compared to optical vortex beams, while FPBs exhibit greater resilience to filamentation when compared to equivalent Gaussian beams. Additionally, we investigate the evolution of the polarization morphologies of these beams and demonstrate that the polarization profile of CVBs remains unchanged, while FPBs undergo notable non-uniform polarization changes during their nonlinear propagation.

The structure of this paper is as follows: we begin with an introduction in Section \ref{sec1}. In Section \ref{sec2}, we provide a detailed account of our model and methodology. Section \ref{sec3} is dedicated to presenting our results and discussions, starting with CVBs in Section \ref{sec3A} and then addressing FPBs in Section \ref{sec3B}. Finally, our conclusions are presented in Section \ref{sec4}.

	\section{Model and method}
	\label{sec2}
	We consider a pulse propagating along $z$-axis, whose electric field is expressed as $ \mathbf{E} (r,t) = \mathrm{Re}\{ \mathbfcal{E}(r,t) e^{i(kz-\omega_0 t)}\} $, where $\mathbfcal{E}(r,t)$ is the slowly varying envelope, $k$ is the wave number, and $\omega_0$ is the central frequency. The field envelope vector can be expanded in terms of an arbitrary orthogonal polarization basis, such as horizontal/vertical or right-/left-circular polarizations. When expressed in the circular polarization basis, the field envelope vector can be written as
	\begin{equation}
		\label{eq1}
		\mathbfcal{E} = \mathcal{E}^+ \hat{\boldsymbol\epsilon_\mathrm{L}}+\mathcal{E}^- \hat{\boldsymbol\epsilon_\mathrm{R}},
	\end{equation}
\noindent where $\hat{\boldsymbol\epsilon_\mathrm{L}}$ and $\hat{\boldsymbol\epsilon_\mathrm{R}}$ are the unit vectors denoting the left-circular and right-circular polarizations, LCP and RCP, respectively; while $\mathcal{E}^+$ and $\mathcal{E}^-$ are the complex amplitudes of the LCP and RCP components, respectively.
	
	For the simulation of the nonlinear propagation, we use the vector-extended version of the standard model of filamentation \cite{KolesikPRE01}, in which the nonlinear propagation equation of the field amplitudes of the two polarization components in an isotropic medium is given by 		
	\begin{equation}
	\begin{aligned}
		\label{eq2}
		\frac{\partial{\mathcal{E}^\pm}}{\partial{z}}=&\frac{i}{2k} \bigtriangleup_{\bot}{\mathcal{E}^\pm} + i \frac{k''}{2} 			\frac{\partial^2 \mathcal{E}^\pm}{\partial t^2}\\
		&-\frac{\sigma}{2} (1+i \omega_0 \tau_c) \rho \mathcal{E}^\pm -  \frac{\beta^{(K)}}{2} |\mathcal{E}|^{2(K-1)} \mathcal{E}^\pm\\
		&+ i \frac{2}{3} k n_2 \left(|{\mathcal{E}^\pm}|^2 +  2 |{\mathcal{E}^\mp}|^2\right) \mathcal{E}^\pm,
	\end{aligned}
	\end{equation}
	where $\bigtriangleup_{\bot}$ stands for transverse Laplacian operator, $|\mathcal{E}|^2 = |\mathcal{E}^+|^2 + |\mathcal{E}^-|^2$, $\mathcal{E}^{\pm}$ are defined in such a way that $|\mathcal{E}^{\pm}|^2$ is given in the units of intensity (e.g., W/cm$^2$), $k'' = \frac{\partial^2 k}{\partial \omega^2}$ is the group velocity dispersion coefficient, $\sigma$ is the cross-section of inverse bremsstrahlung, $\tau_c$ is the electron collision time, $\beta^{(K)}$ is the $K$-photon absorption coefficient, $\rho$ is the electron density and $n_2$ is the nonlinear Kerr index. The first and second terms on the right-hand side stand for diffraction and dispersion, respectively, and the last five terms represent absorption by plasma, plasma defocusing, multiphoton absorption (MPA), and instantaneous Kerr effect, respectively. 
	
	For an arbitrary input polarization, the instantaneous Kerr effect is composed of two contributions: a ``self"-Kerr term, and a ``cross"-Kerr term. The self-Kerr term depends on the intensity of the identical polarization component, while the cross-Kerr term scales with the intensity of the orthogonal polarization component. Additionally, in the case of non-resonant nonlinear electronic response, the cross-Kerr term is twice as effective as the self-Kerr term in isotropic media \cite{BoydNLO20,KolesikPRE01, RostamiOE15}. In fact, this is due to the symmetry properties of the third-order susceptibility tensor in isotropic media (as discussed in detail in Section I of the Supplementary Material \cite{SuppMat2024}). 
	
	Self-focusing due to optical Kerr nonlinearity can balance the diffraction only if the input beam power ($P_\mathrm{{in}}$) exceeds the critical power of self-focusing, $P_{\mathrm{cr}} \simeq \lambda^2_0/2\pi n_0 n_2$, where $\lambda_0$ is the central wavelength, and $n_0$ is the refractive index of the propagation medium  \cite{CouaironPhysRep07}.

	Moreover, to account for the effect of plasma, Eq. (\ref{eq2}) must be simultaneously solved with the rate equation of the electron density 
	\begin{equation}
		\frac{\partial{\rho}}{\partial{t}} = \sigma_K |\mathcal{E}|^{2K} (\rho_{at} - \rho) + \frac{\sigma}{U_i} \rho |\mathcal{E}|^2 - a \rho^2,
		\label{eq3}
	\end{equation}
	where $\sigma_k$ is the cross-section for multiphoton ionization (MPI), $\rho_\mathrm{at}$ is the density of neutral atoms, $\sigma$ is the cross-section for inverse bremsstrahlung, ${U}_\mathrm{i}$ is the ionization potential of the medium, and $a$ is the recombination rate. The first term on the right-hand side stands for multiphoton ionization, the second term represents avalanche ionization, and the last term denotes plasma recombination. For near-IR laser pulses with a pulse duration of a few tens of femtoseconds, propagating in air, the last two terms in Eq. (\ref{eq3}) can be neglected \cite{KolesikPRE01}.
	
	In order to investigate the effect of polarization on different features of filamentation, we  consider the input beam to have a general form of a Poincar\'e beam whose field envelope is constructed by a vector superposition of two optical angular momentum (OAM)-carrying spatial transverse eigenmodes with different topological charges and orthogonal circular polarizations
	\cite{GibsonPRA18,BouchardPRL16}
	\begin{equation}
				\mathbfcal{E} = \cos(\gamma) \mathrm{LG}_{m,l} \hat{\boldsymbol\epsilon_L}
				+\sin(\gamma) e^{i\beta} \mathrm{LG}_{m',l'} \hat{\boldsymbol\epsilon_R}.
		\label{eq4}
	\end{equation}
	
	The complex amplitudes of the LCP and RCP components are in the form of the Laguerre-Gaussian modes,  
	$\mathrm{LG}_{m,l}$ \cite{Kobolov07} where $m$ and $l$ are the radial and azimuthal indices, respectively. The amplitudes of the polarization components are determined by parameter $\gamma$, while their phase difference is denoted by $\beta$. The temporal envelope of the pulse is considered to have a Gaussian shape in the form of $\exp{[-(t/t_p)^2]}$ with a width of $t_p$. Most generally, the LCP and RCP components may have different azimuthal and radial indices with arbitrary relative amplitudes and phases.
	
	For different values of $\gamma$, $\beta$, $m$, $m'$, $l$, and $l'$ various polarization morphologies can be generated. We generate CVBs with azimuthal, radial, and spiral polarization profiles along with two low-order FPBs with ``star'', and ``lemon'' polarization topologies for different values of the parameters $\gamma$, $\beta$, $l'$ and $l$, all with $m=m'=0$. In our numerical investigations, we consider an initial pulse width of $t_{p}=35$ fs and a beam width of $w_{0}=0.5\ \text{mm}$ (both, half-widths at $e^{-1}$ of the maximum amplitude), and a central wavelength of $\lambda_{0}=800$ nm. The propagation medium is air for which $P_\mathrm{cr}= 3.2$ GW, for a linearly polarized Gaussian beam. The values of the physical parameters used in the model are given in Table \ref{tab1} \cite{ParisPRA12}.
	\begin{table}[h]
	\caption{Physical parameters for $\lambda_0 = 800$ nm and a pulse width of 35 fs used in our simulations.}
		\label{tab1}
		\begin{ruledtabular}
			\begin{tabular}{ll}
				\textrm{Quantity}&
				\textrm{Value and unit}\\
				\colrule
				$n_0$	& $1$ \\ 
				$n_2$	& $3.2 \times 10^{-19} \mathrm{cm^2/W}$ \\ 
				$k''$	& $0.2 \ \mathrm{fs^2/cm}$ \\ 
				$K$	& $8$\\
				$\beta^{(K)}$	& $4 \times 10^{-95} \mathrm{cm^{13}/W^7}$ \\ 
				$\sigma_K$	& $4 \times 10^{-96} \mathrm{s^{-1} cm^{16}/W^8}$ \\ 
				$\sigma$	& $5.5 \times 10^{-20} \mathrm{cm^{2}} $\\  
				$\rho_{at}$	& $5 \times 10^{18} \mathrm{cm^{-3}}$\\ 
				$\tau_{c}$	& 350 fs\\
			\end{tabular}
		\end{ruledtabular}
\end{table}
	
	For numerically solving the nonlinear propagation equation, Eq. (\ref{eq2}), we use alternate direction implicit (ADI) scheme based on the Crank-Nicolson method, which is an implicit unconditionally stable finite difference method for solving partial differential equations in the form of the paraxial beam propagation equation \cite{CouaironEPJ11}. The nonlinear terms in Eq. (\ref{eq2}) were implemented in the numerical simulation using the second-order Adams-Bashforth scheme \cite{CouaironEPJ11}. The use of the ADI scheme enables numerical simulations in four dimensions ($x,y,z,t$), without assuming any sort of symmetry.
	
	Moreover, for nearly realistic simulations of the nonlinear propagations in the presence of turbulence and imperfections we added uniform amplitude noise with a correlation length of $0.15 w_0$, and two different levels of 1\% and 10\% of the initial peak amplitude of the beams. The results of the simulations with 1\% noise amplitude will be presented in the following, while the results of the simulations with 10\% noise amplitude are presented in the Supplementary Material \cite{SuppMat2024}.

\section{Results and discussion}
	\label{sec3}
	\subsection{Filamentation of cylindrical vector beams}
	\label{sec3A}
	Different CVBs can be generated by setting $m'=m=0$, and $l'=-l =1$ in Eq. (\ref{eq4}), such that the superposition of the orthogonally circular polarized LG components carries zero net OAM, and cylindrically symmetric polarization morphology is determined by the relative amplitude (i.e., through the parameter $\gamma$) and the phase difference ($\beta$) of the constituting LCP and RCP LG components. Here, we compare the filamentation of LCP LG$_{0,1}$ beam, and three CVBs with different parameters and polarization morphologies: 
	
	\begin{enumerate}[label=(\alph*),itemsep=0mm]
		\item ($\gamma=0$): LCP LG$_{0,1}$ beam
		\item ($\gamma=\pi/4$, $\beta=\pi$): Azimuthal CVB
		\item ($\gamma=\pi/4$, $\beta=0$): Radial CVB
		\item ($\gamma=\pi/4$, $\beta=\pi/2$): Spiral CVB.
	\end{enumerate}
	
\begin{figure*}[t]
	\includegraphics[width=0.95\textwidth]{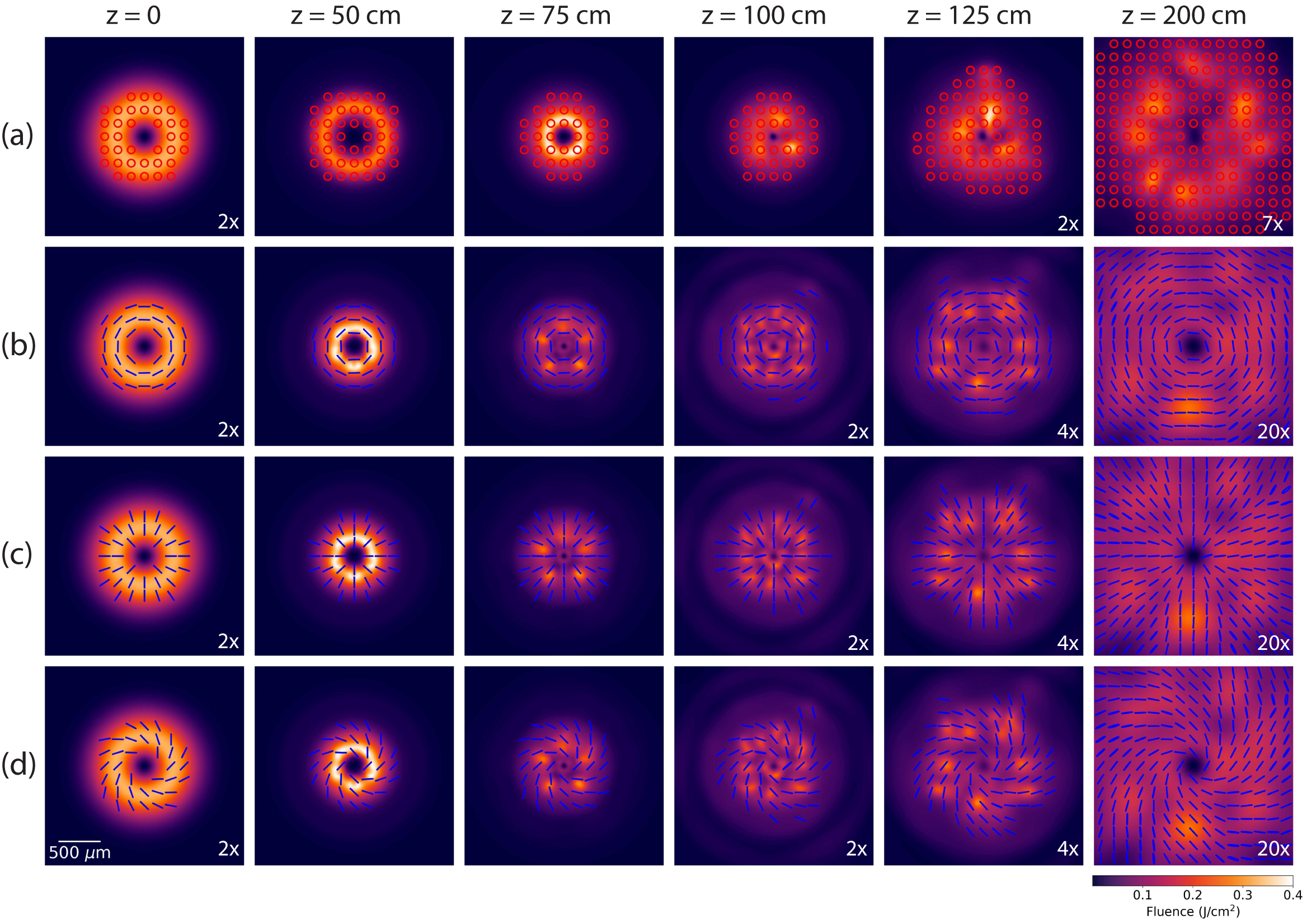} 
		\caption{Transverse fluence and polarization profiles of the beams at different propagation distance: (a) LCP LG$_{0,1}$ beam, (b) azimuthal CVB, (c) radial CVB, and (d) spiral CVB. To maintain consistency in the colorbar across all fluence maps, some maps are scaled by specific factors indicated in the bottom-right corner of each map. In the polarization profiles, red represents left-circular polarization (LCP), while blue indicates linear polarization.}
		\label{fig1}
\end{figure*}	
	The transverse fluence profiles of the beams and their polarization morphologies at the initial propagation distance ($z=0$) are shown in Fig. \ref{fig1} (leftmost column). The polarization profiles on arbitrary transverse planes are mapped using the Stokes parameters \cite{GibsonPRA18, DennisProgOpt09}
	\begin{equation}
	\begin{aligned}
	S_{0}&=\mathcal{I}=|\mathcal{E}^{-}|^{2}+|\mathcal{E}^{+}|^{2}, &
	S_{1}&=2\ \text{Re}\{\mathcal{E}^{-^{*}}\mathcal{E}^{+}\},\\
	S_{2}&=2\ \text{Im}\{\mathcal{E}^{-^{*}}\mathcal{E}^{+}\}, &
	S_{3}&=|\mathcal{E}^{-}|^{2}-|\mathcal{E}^{+}|^{2},
	\end{aligned}
	\label{eq5}
	\end{equation}
\noindent and the local polarization ellipticity, $\chi$, and polarization orientation, $\psi$:
\begin{align}
	\chi&=\frac{1}{2}\sin^{-1}\left(\frac{S_{3}}{S_{0}}\right), \label{eq6}\\
	\psi&=\frac{1}{2}\tan^{-1}\left(\frac{S_{2}}{S_{1}}\right). \label{eq7}
\end{align}
The transverse fluence maps are calculated by time integration of the intensity profiles at each propagation distance, $\mathcal{F}(x,y,z) = \int_{-\infty}^{\infty} \mathcal{I}(x,y,z,t) dt$, with $x$, and $y$ representing the transverse Cartesian coordinates.
	
	As mentioned earlier, filamentation can occur only if the input beam power exceeds the critical power, $P_\mathrm{cr}$. The critical power depends on the input beam profile and the values given for $P_\mathrm{cr}$ in the literature are usually associated with either Gaussian or flat-top beams, while it is shown that the optical vortex beams (i.e., OAM-carrying optical beams, such as the LG beams) are more resilient to self-focusing and collapse \cite{VuongPRL06, PolynkinPRL13}, and the critical power of self-focusing for optical vortex beams depends on their topological charge, $l$. Particularly, for $l=\pm1$ it is shown that the critical power is 3.85 times higher than that of a corresponding Gaussian beam, $P_\mathrm{cr}^{(1)}=3.85P_\mathrm{cr}$ \cite{VuongPRL06}. Therefore, in this investigation, the input power of all four beams is set to 39 GW, corresponding to 3.16$P_\mathrm{cr}^{(1)}$, to ensure that the polarization components in the form of LG$_{0,\pm1}$ modes can form filaments. 	

	Equation (\ref{eq2}) intuitively implies that during the nonlinear propagation of the pulse, the spatiotemporal profiles as well as the polarization profile of the beams may evolve because of the combined actions of different linear and nonlinear effects. Starting from the intensity evolution of the beams during propagation in the filamentation regime, Fig. \ref{fig1} illustrates the transverse fluence and polarization profiles of the LCP LG$_{0,1}$ beam and the three CVBs at different propagation distances, with identical initial noise profile (with a level of 1\% of the peak initial amplitude and correlation length of 0.15$w_0$). In order to use the same fluence representation limits for each beam at all propagation distances, some of the fluence maps are multiplied in constant values indicated on the right bottom corner of the profiles. 
		
		Although the maximum fluence is comparable for all cases, the onset position of the filamentation, filamentation length, and number of the filamentary structures are different in some cases. The LCP LG$_{0,1}$ beam (Fig. \ref{fig1}(a)) begins filamentary propagation from the farthest distance, and it generates four transverse filamentary structure, after the nonlinear focus (Fig. \ref{fig1}(a) at $z = 100$ cm). This is consistent with the predicted number of filamentary structures for a LG beam with $|l|=1$ \cite{VuongPRL06}, considering the fact that the critical power for a circularly polarized beam is 1.5 times the critical power of the same beam with linear polarization \cite{KolesikPRE01}.
	Moreover, it is clearly seen that the azimuthal, radial, and spiral CVBs (Fig. \ref{fig1}(b-d), respectively) have an identical propagation profile. 
	Filamentation of the CVBs initiates from a shorter propagation distance and they initially generate five filaments (as illustrated in Fig. \ref{fig1}(b-d) at $z=75$ cm). Interestingly, the number of intense transverse spots increases at longer propagation distances. We attribute this effect to the fact that the filaments generated at shorter propagation distances, close to the nonlinear focus, undergo nonlinear losses while the small transverse intensity fluctuations (due to the initial noise) are amplified due to the strong Kerr effect, and appear as hot spots in the transverse fluence maps, at longer propagation distances. The transverse sizes of the intense spots close to the nonlinear focus are comparable for all beams ($\sim$$180\ \mu\text{m}$, FWHM).
		 
	Moreover, a comparison of the polarization profile of each beam with its respective initial polarization profile, shown in Fig. \ref{fig1}(a-d), reveals that the polarization distributions of the beams are preserved upon filamentation because of the absence of RCP polarization component in the case of the LCP LG$_{0,1}$ beam, and the balance between the two polarization components in the case of the CVBs.
	 
\begin{figure}[t]
		\includegraphics[width=0.48\textwidth]{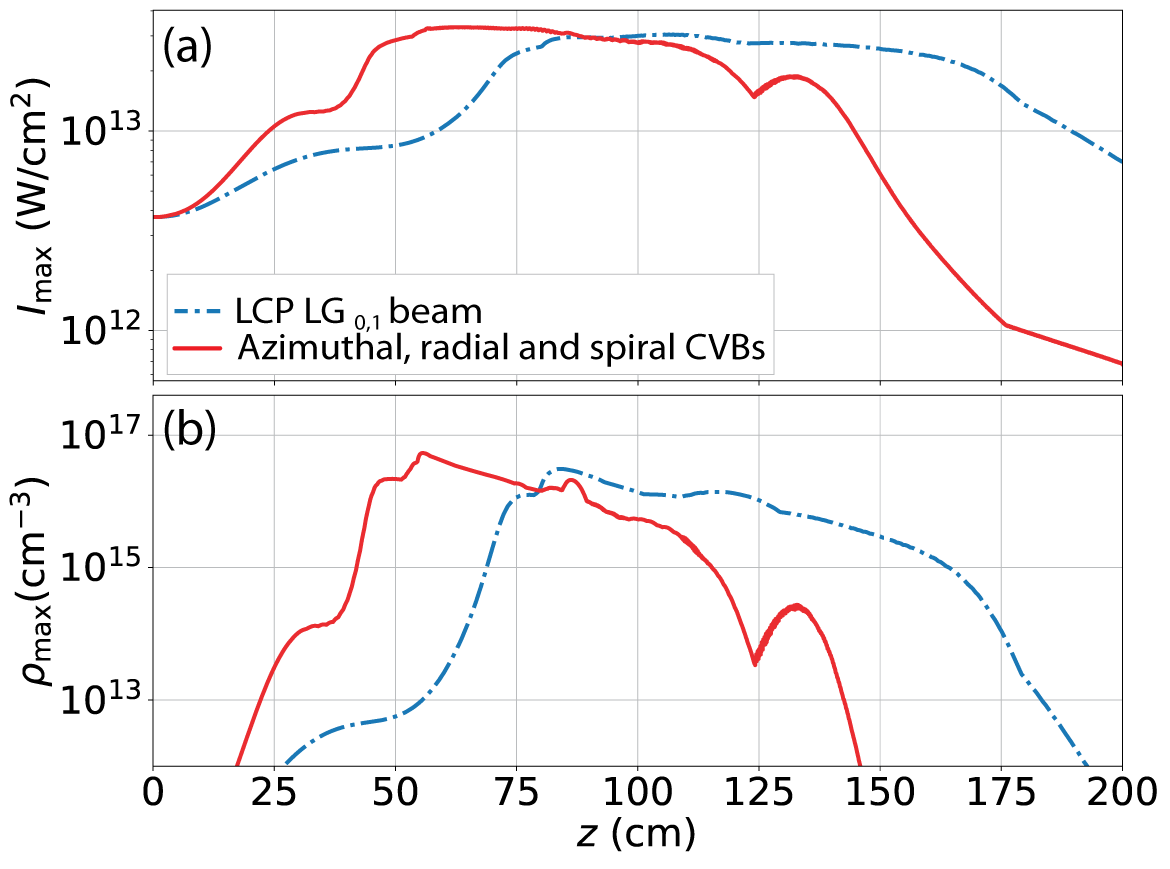} 
		\caption{Evolution of maximum intensity (a), and maximum electron density (b) during the filamentation of LCP LG$_{0,1}$ beam and the three CVBs. Dash-dotted (blue) curve: LCP LG$_{0,1}$; solid (red) curve: azimuthal, radial, and spiral CVBs.}
		\label{fig2}
\end{figure}
In addition to the transverse fluence profiles, shown in Fig. \ref{fig1}, further quantitative analysis can be performed by investigating the evolution of the maximum intensity and the maximum electron density values achieved throughout the spatiotemporal extensions of the pulse as a function of propagation distances (Fig. \ref{fig2}(a-b), respectively). The peak values of the maximum intensity ($I_{\mathrm{max}}$), for the LCP LG$_{0,1}$ beam and the CVBs are (3.1 and 3.3)$\times10^{13}$ W/cm$^2$,  respectively. Although these values are comparable to some extent, their differences indicate distinctions in the filamentation of the beams, since the nonlinear terms in the governing equations of the propagation (i.e., Eqs. (\ref{eq2}-\ref{eq3})) scale with the powers of the pulse intensity. These graphs also show that the filamentation length and its starting position differ for different cases. By defining the filamentation length as the axial distance over which the intensity remains above 2$\times 10^{13}$ W/cm$^2$, we find that the filamentary propagation of the LCP LG$_{0,1}$ beam begins from $z=71$ cm, and the filamentation length is 99 cm, while for the CVBs, filamentation starts from and $z=43$ cm, with a length of 76 cm.

Additionally, the maximum electron density ($\rho_\mathrm{max}$) graphs, shown in Fig. \ref{fig2}(b), reveal that the CVBs yield higher peak electron density than LCP LG$_{0,1}$ beam. The peak maximum electron densities achieved for the LG$_{0,1}$ beam, and the CVBs are (3.1, and 5.4)$\times$10$^{16}$ cm$^{-3}$, respectively.

	It is interesting to note that although all four beams had the same initial power and similar intensity profiles, their filamentary propagation is different. The difference in the filamentation of the beams can be interpreted in terms of the distinct action of the optical Kerr effect on the evolution of the intensity of the beams that is consequently mapped onto the higher-order nonlinear effects, as the following: In the case of the LCP LG$_{0,1}$ beam, the cross-Kerr effect totally vanishes since the RCP component does not exist and the whole input power is carried by the LCP component. Hence, the beam is nonlinearly focused solely due to the self-Kerr effect, and MPA kicks in as the local intensity builds up to produce plasma. Also, an increase in the plasma density leads to plasma defocusing and absorption of the pulse energy by the plasma through inverse bremsstrahlung, though MPA is more effective as it grows with a higher power of intensity. In this case, the farther onset position of the filamentation, longer filamentation length, and smaller number of the filaments are due to the weaker nonlinear focusing, and hence lower nonlinear losses. Conversely, for the three CVBs, since both RCP and LCP components have equal amplitudes, both the self-, and cross-Kerr effects contribute to the nonlinear focusing of each beam, and the phase difference between the components, which determines the polarization morphology, has no effect. Therefore, the nonlinear propagation profiles of all three beams are identical. The equal contributions of the polarization components in the nonlinear focusing of the whole beam, in these cases, result in filamentation onset at a shorter propagation distance, higher number of filaments, higher peak intensities, and consequently higher electron densities originating from MPI, which scales with the $K$$^{\text{th}}$ power of the intensity. Furthermore, higher peak intensities due to stronger nonlinear focusing also lead to more energy loss through MPA for the CVBs, and consequently, the filamentation length of the CVBs is shorter than that of the LCP LG$_{0,1}$ beam. 
	 
	We also investigated the influence of higher noise level (10\% of the initial peak amplitude) on the nonlinear propagation of both the radial CVB and the LCP LG$_{0,1}$ beam. The fluence maps of the beams along with their polarization profiles at different propagation distances, and also the evolution of the maximum intensity and electron density as a function of the propagation distance are illustrated in Supplementary Material (Fig. S. 1, and Fig. S. 2, respectively)\cite{SuppMat2024}. It is shown that higher input noise level does not affect the number, and the size of the filamentary structures in either case. Additionally, these data also show that the peak intensity and the maximum electron density do not change significantly by increasing the noise level.
 	
\subsection{Filamentation of Full Poinca\'e beams}
\label{sec3B}
\begin{figure}[t]
		\includegraphics[width=0.49\textwidth]{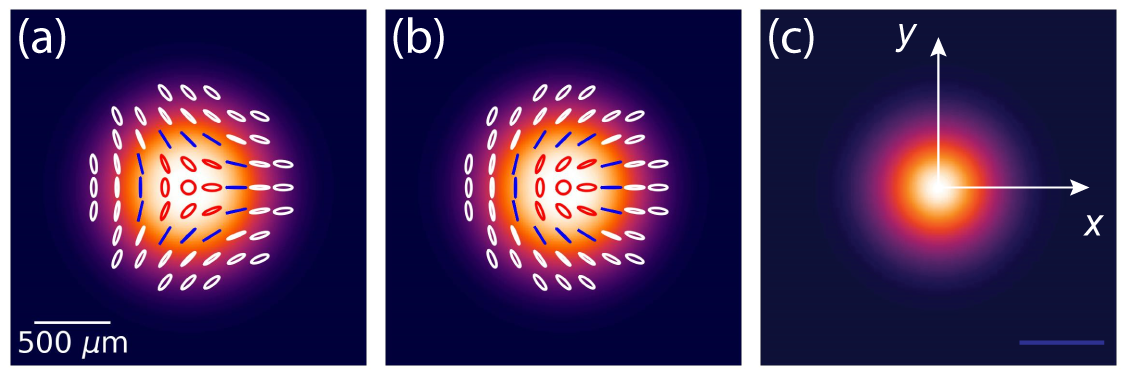}
		\caption{Transverse fluence, and polarization profiles of the star FPB (a), lemon FPB (b), and linearly-polarized Gaussian beam (c), at the initial propagation distance ($z=0$). In the polarization profiles, red, and white colors represent left-, and right-handed polarization states respectively; while blue color indicates linear polarization.}
		\label{fig3}
\end{figure}
	We also examine the filamentation of the lowest-order Full Poincar\'e beams (FPBs), with polarization topologies known as ``star" and ``lemon" \cite{BouchardPRL16}. 
	The star polarization topology can be generated by setting $\gamma = \pi/4$, $\beta = 0$,  $m'=m=0$, $l=0$, and $l^\prime = + 1$, while the lemon polarization topology is generated by setting $l^\prime = - 1$, with otherwise the same parameters. Figures \ref{fig3}(a, b) displays the transverse fluence profiles of the star and lemon FPBs, respectively, along with their corresponding polarization topologies at $z=0$. The intensity profiles of both beams closely resemble a Gaussian distribution. Therefore, we compare the filamentary propagation of both beams with that of a linearly polarized (LP) Gaussian beam with equal input power (i.e. $17\ \mathrm{GW}\simeq5.3$$P_\mathrm{cr}$), and the same width (i.e., $w_{0}=0.5$ mm), whose fluence profile is depicted in Fig. \ref{fig3}(c). Similarly, an identical initial uniform amplitude noise profile with a 1\% amplitude, and a correlation length of 0.15$w_0$ is added to the initial field amplitudes of the three beams.
 	
\begin{figure}[t]
	\includegraphics[width=0.48\textwidth]{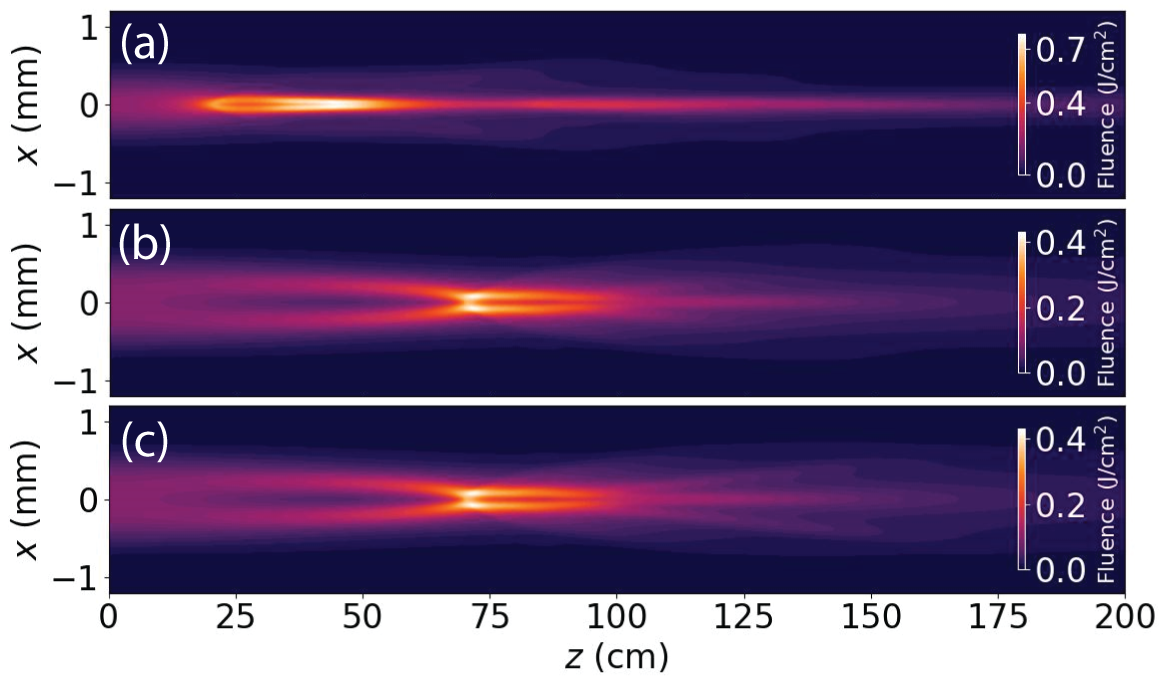}
	\caption{Propagation fluence cross-sections of the beams during filamentation: (a) LP Gaussian beam; (b) and (c) Full Poinca\'e beams with the star and lemon polarization morphologies, respectively.}
	\label{fig4} 
\end{figure}

	The $x$-$z$ cross-sections of the fluence profiles, $\mathcal{F}(x,z)$, of the nonlinear propagation of the LP Gaussian beam, and star, and lemon FPBs are presented in Fig. \ref{fig4}(a-c), respectively. Obviously, both star and lemon Poincar\'e beams form a transverse ring-like filamentary structure in their filamentation range (from $\sim$68 cm to $\sim$135 cm, as it will be discussed later). Moreover, the maximum fluences of the star and lemon Poincar\'e beams during their filamentation are identical and about two times smaller than that of the LP Gaussian beam. Furthermore, the diameter of the filament for the LP Gaussian beam is about 150 $\mu$m, while the thickness (both FWHM) of the filamentary ring-like structure of the Full Poincar\'e beams is slightly bigger, about $\sim$190 $\mu$m.
	
\begin{figure}[t]
		\includegraphics[width=0.48\textwidth]{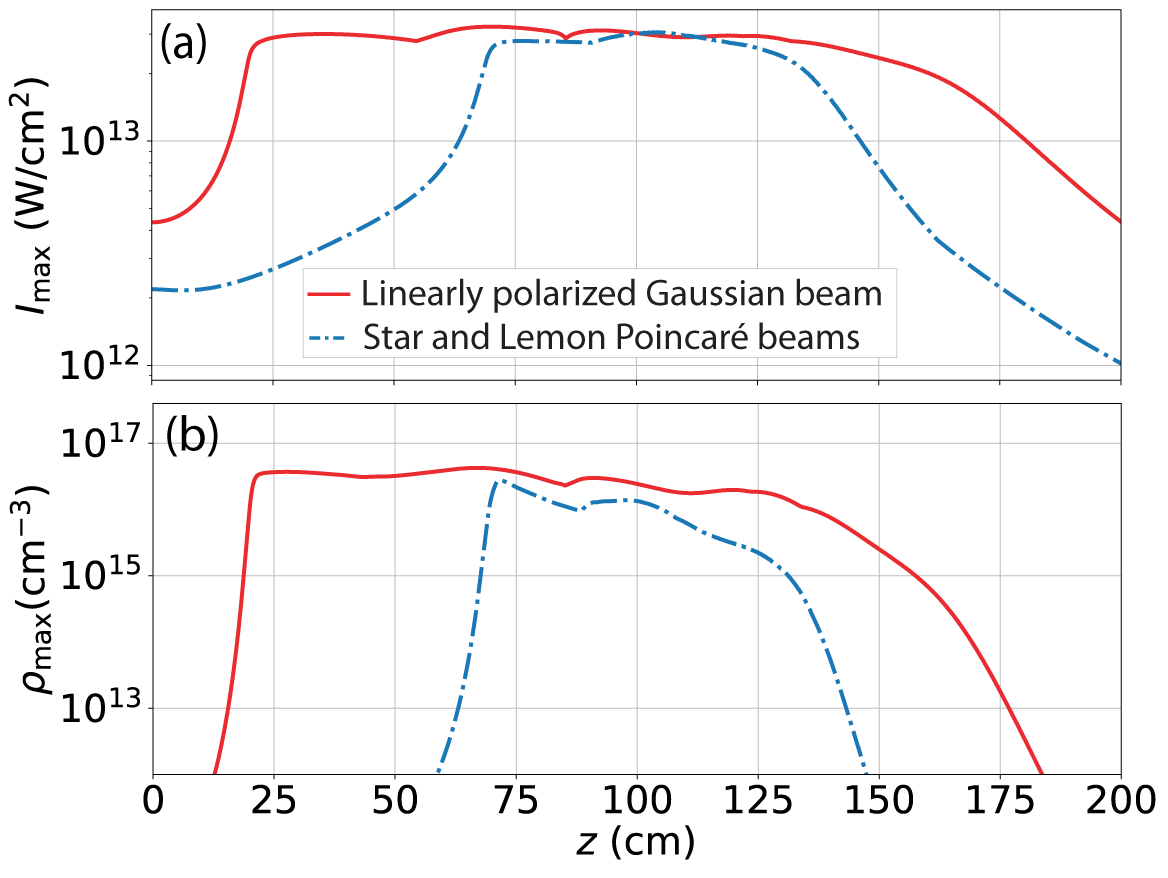} 
		\caption{Evolution of the maximum intensity (a), and maximum electron density (b) during the filamentation of the LP Gaussian beam, and Poincar\'e beams with star and lemon polarization topologies.}
		\label{fig5}
\end{figure}

The evolution of the maximum intensity and the maximum electron density during the nonlinear propagation of the three beams are shown in Fig. \ref{fig5}(a-b), respectively. For all three beams, the maximum intensity is clamped at $\sim$3$\times 10^{13}$ W/cm$^2$, confirming that the filamentation occurs for all beams. For the LP Gaussian beam, filamentation begins at $z=19$ cm with a length of 141 cm. Conversely, for the Poincar\'e beams filamentation onsets at $z=68$ cm and the filamentation length is 67 cm, about 2 times shorter than that of the LP Gaussian beam. Moreover, the maximum electron density during the filamentation of the Gaussian beam reaches $4.3\times10^{16}\ \text{cm}^{-3}$, and a lower value of $2.8\times10^{16}\ \text{cm}^{-3}$ for both FPBs.

The differences in the filamentation of the LP Gaussian beam and FPBs can be understood based on a similar description presented for the filamentation of CVBs in Section \ref{sec3A}, with certain differences due to distinct mode structures of the LCP and RCP components of the FPBs, and their interactions during the nonlinear propagation. The LP Gaussian beam can be considered to be composed of two Gaussian beams with orthogonal circular polarizations, but otherwise identical properties. Therefore, the nonlinear propagation of each circular polarization component of the LP Gaussian beam is similarly affected by the orthogonal polarization component through the cross-Kerr effect. On the other hand, for the star and lemon Poincar\'e beams, the total input power of the beam is equally divided between their LCP Gaussian and RCP LG$_{0,\pm1}$ components. The power carried by the LCP Gaussian component corresponds to 2.66$P_{\text{cr}}$, while the power contained in the RCP LG$_{0,\pm1}$ component corresponds to 0.7$P_{\mathrm{cr}}^{(1)}$. Moreover, at $z=0$, the peak intensity of the LCP Gaussian mode is 2.72 times higher than that of the RCP LG$_{0,\pm1}$ modes. Therefore, the LG modes undergo nonlinear focusing solely due to the cross-Kerr effect, while the RCP Gaussian component undergoes nonlinear focusing both due to the self-, and cross-Kerr effects although the self-Kerr focusing is slightly stronger. The resilience of the LG modes of the FPBs to self-focusing results in a weaker nonlinear focusing of the whole FPB and consequently the filamentation of the FPBs begins from a longer propagation distance and with larger transverse dimensions of the filamentary structures.

\begin{figure}[t!]
	\includegraphics[width=0.48\textwidth]{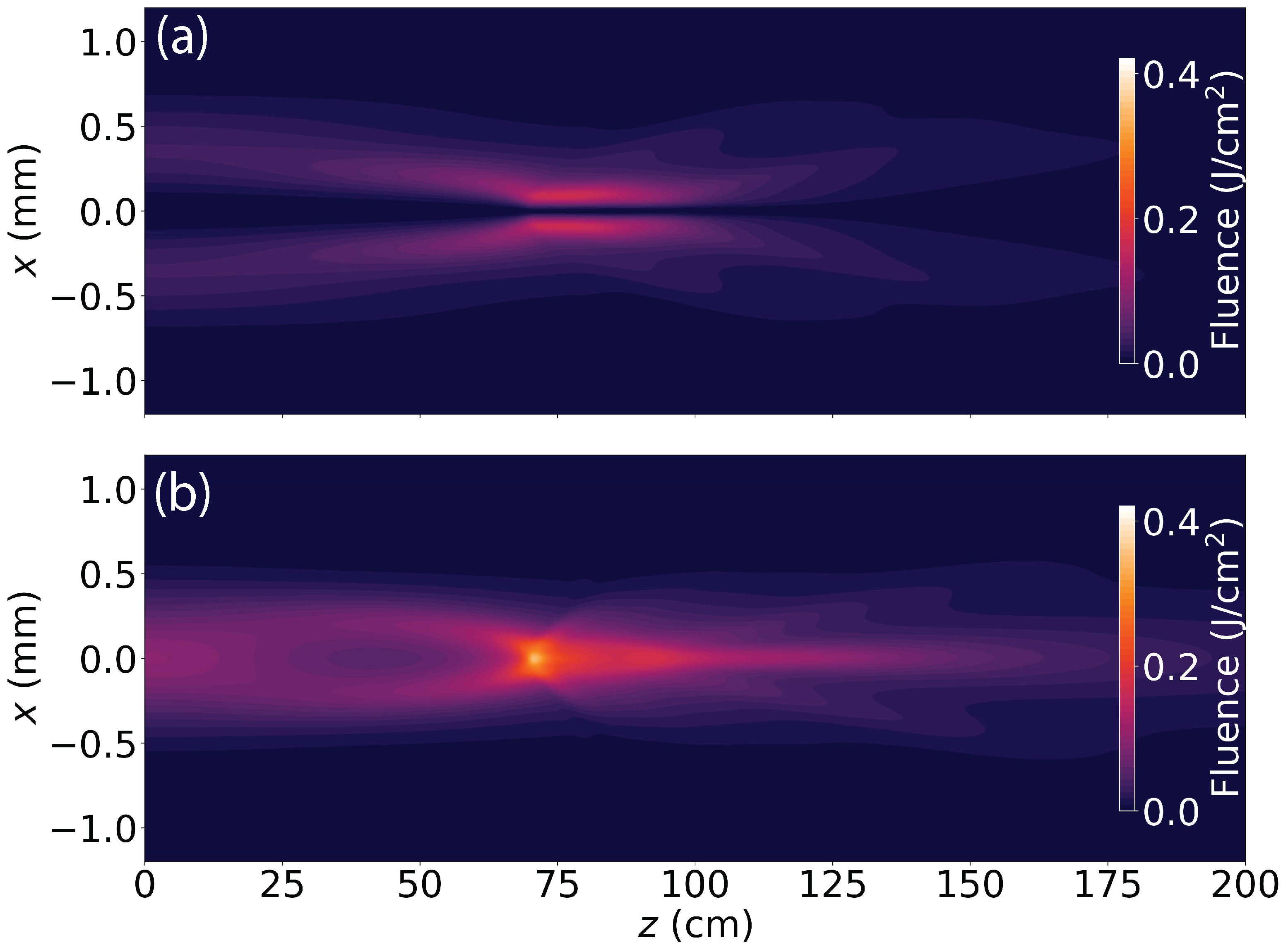} 
	\caption{Propagation fluence cross-sections of the orthogonal polarization modes of the lemon FPB during filamentation: (a) RCP LG$_{0,-1}$ mode, and (b) LCP Gaussian mode.}
	\label{fig6}
\end{figure}

It is also important to note that in the case of the FPBs, different mode shapes of the two polarization components lead to mode reshaping during the nonlinear propagation under the action of the Kerr effect. The fluence propagation cross-section profiles of the RCP LG mode and LCP Gaussian mode of the lemon FPB during its nonlinear propagation are presented in Fig. \ref{fig6}(a, b), respectively. The cross-Kerr effect of the LG mode acting on the LCP Gaussian mode reshapes the beam to a hollow beam at the initial stages of the propagation, before the onset of filamentation (Fig. \ref{fig6}(b)). On the other hand, the cross-Kerr effect of the LCP Gaussian mode acting on the LG mode results in the nonlinear focusing of the LG mode that leads to an increase in the nonlinear refractive index over the doughnut-shaped profile of the LG mode. 
As a result, maximum fluence over the RCP LG mode reaches a maximum value of $\sim$$0.2$ J/cm$^{2}$ (Fig. \ref{fig6}(a)), while that of the LCP Gaussian mode exceeds 0.4 J/cm$^{2}$ (Fig. \ref{fig6}(b)). The factor two difference between the two peak fluences, implies a similar difference between the peak intensities, and based on Eq.(\ref{eq3}) it can be inferred that multiphoton absorption is more effective for the LCP Gaussian mode than the RCP LG mode, and therefore the plasma generation is almost solely performed by the LCP Gaussian mode. The slight refractive index increase over the LG mode due to the nonlinear focusing, in the absence of effective plasma density, leads to the  reshaping of the Gaussian mode. This reshaping creates a central spot surrounded by an intense structures that mimic the profile of the LG mode. The transverse fluence maps of the two constituent modes of both FPBs during their filamentation with a step size of 5 cm is shown in Supplementary Video 1 \cite{SuppMat2024}.

On the other hand, comparing the filamentation of the LP Gaussian beam (Fig. \ref{fig4}(a)) with that of the FPBs (Fig. \ref{fig4}(b-c)), an almost two fold difference between the maximum fluences of the LP Gaussian beam and that of the Poincar\'e beams can be explained based on the fact that in the former case, both RCP and LCP components contribute in filamentation, while in the latter only one component (i.e., the LCP Gaussian component) contributes in filamentary propagation. Furthermore, the power carried by the Gaussian mode of FPBs is twice lower than that of the LP Gaussian beam and thus the filamentation length of the FPBs is shorter than that of the LP Gaussian beam by a comparable factor. This is in agreement with the estimated relation between the filamentation length and the pulse energy \cite{CouaironAPB03}.

\begin{figure*}[ht!]
		\includegraphics[width=1\textwidth]{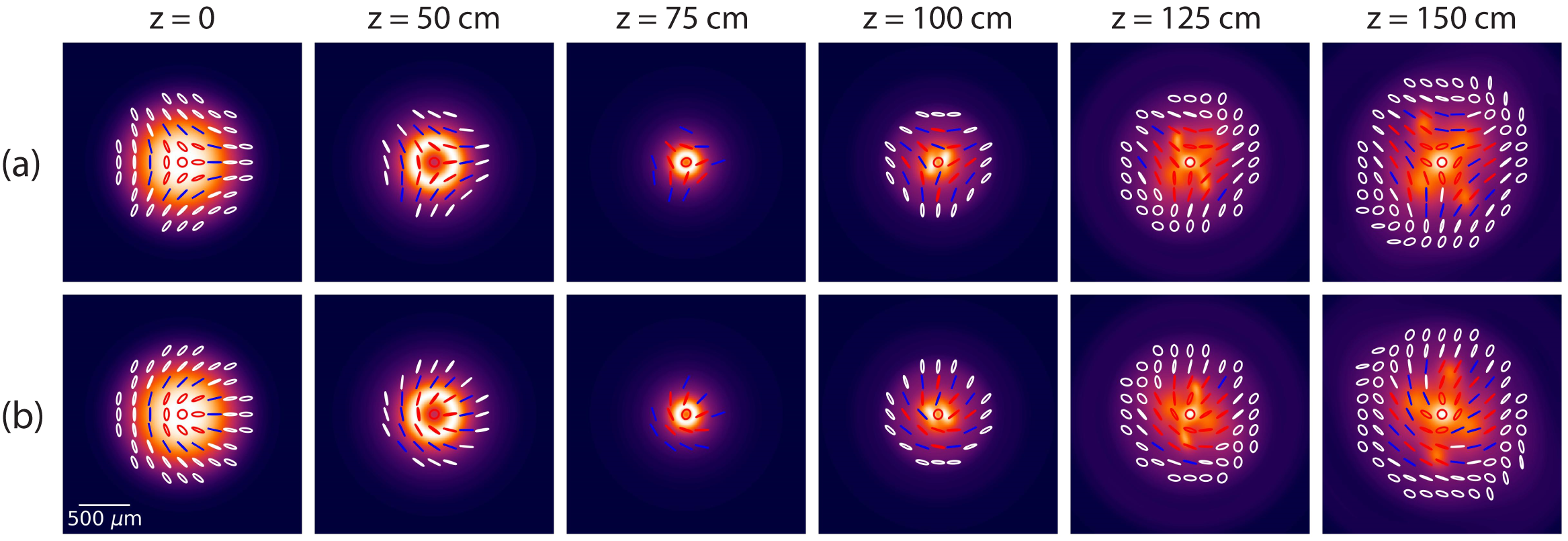} 
		\caption{Evolution of the transverse fluence and polarization topologies of the star (a), and lemon (b) Full Poincar\'e beams during their filamentation. In the polarization profiles, red, and white colors represent left-, and right-handed polarization states respectively; while blue color indicates linear polarization.}
		\label{fig7}
\end{figure*}
	
	We also investigate the evolution of the polarization profiles of the star and lemon FPBs during their filamentation. The transverse fluence and polarization profiles of the star and lemon beams at six propagation distances are shown in Fig. \ref{fig7}(a) and Fig. \ref{fig7}(b), respectively. The profiles are also illustrated in Supplementary Video 2 \cite{SuppMat2024}, with a finer step size of 5 cm. Additionally, the transverse fluence and polarization profiles at several propagation distances during the filamentation of the lemon beam with a higher noise level (e.g., 10\% of the initial peak amplitude) are shown in Fig. S. 3, and corresponding maximum intensity and electron density are shown in Fig. S. 4 \cite{SuppMat2024}.
	 	
\begin{figure}[t]
	\includegraphics[width=0.49\textwidth]{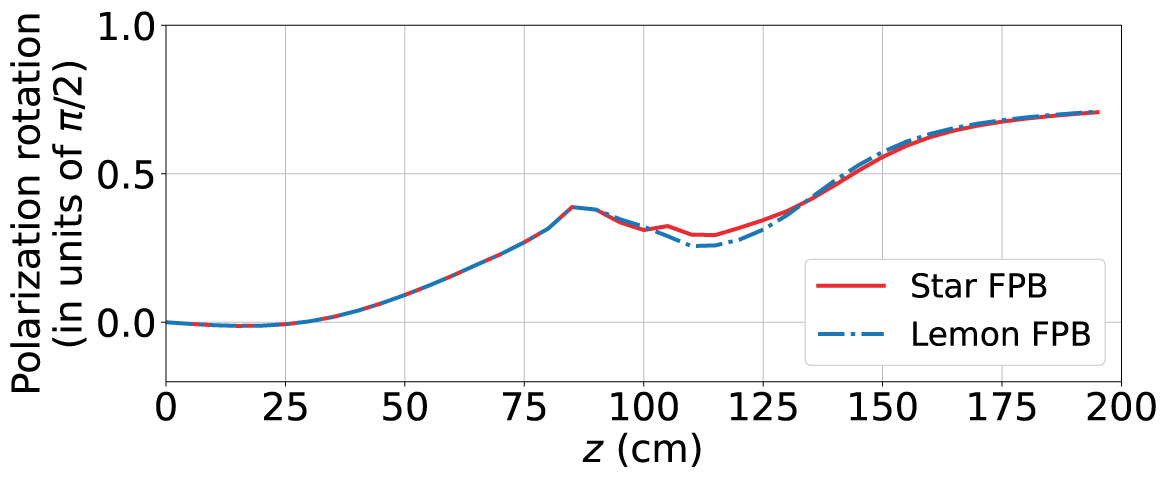} 
	\caption{Average polarization rotation over the polarization states located on the most intense parts of the star and lemon FPBs as a function of the propagation distance.}
	\label{fig8}
\end{figure}
	 	 
	For both Poincar\'e beams, comparing the orientation of the polarization states at various propagation distances with those at $z=0$, reveals that the polarization states do not rotate uniformly across the transverse planes. This behavior contrasts with linear propagation, where a uniform $\pi/2$ counterclockwise  rotation of each polarization state is expected as the beam (either star or lemon) travels over several Rayleigh ranges from one side of the waist plane to the other, due to the Gouy phase difference between the Gaussian and LG components \cite{GibsonPRA18}. More specifically, during the linear propagation of our FPBs with $w_0 = 500\ \mu$m (corresponding to a Rayleigh length of $z_R\simeq98.2$ cm), a counterclockwise  rotation of about 0.178$\pi$ is expected to occur from $z=0$ to $z= 2$ m. For a quantitative assessment of the polarization rotation during filamentation of the FPBs, we measured the orientation of each polarization state and calculated the rotation of each state during the propagation. The average value of the rotation over polarization states located on the most intense parts of the star and lemon FPBs (at a radius $\sim$$150\ \mu$m), as a function of the propagation distance, are shown in Fig. \ref{fig8}. Both Fig. \ref{fig7} and Fig. \ref{fig8} (and the Supplementary Video 2 \cite{SuppMat2024}) demonstrate that the polarization states in the most intense regions of the star and lemon FPBs consistently rotate in the same direction as the propagation distance increases. Figure \ref{fig8} illustrates that at shorter propagation distances, the polarization states of both beams exhibit a slight clockwise (negative) rotation up to approximately $z\simeq 15$ cm. Following this, the polarization states rotate counterclockwise until  $z=85$ cm. At this point, the rate of rotation decreases and reverses direction, transitioning back to a clockwise rotation until $z= 110$ cm. Beyond this position, the polarization states resume a counterclockwise rotation, continuing through the remainder of the observed propagation. 
		
	In addition to the nonuniform polarization rotation, it is noteworthy that the ellipticity of the polarization states, represented by 
	$\chi$ in Eq. (\ref{eq6}), increases in the intense regions of the beams (excluding the beam axis) during the initial stages of propagation. This leads to the development of nearly linear polarization states around the nonlinear focus, between $z=70$ cm to $z=90$ cm. Beyond this propagation range, right-handed elliptical polarization states begin to re-emerge at the outer regions of the beam, and as the propagation distance increases further, left-handed polarization states also appear near the beam axis.
	
	The complex and nonuniform dynamics of polarization reshaping and rotation in the nonlinear propagation regime can be better understood by examining the intricate mode reshaping process occurring during the filamentation of the beams, as illustrated in Fig. \ref{fig7} and in Supplementary Video 2 \cite{SuppMat2024}. Unlike the linear propagation regime—where polarization rotation is solely driven by the diffraction of the constituent modes and their associated Gouy phase difference—the nonlinear regime involves significant mode reshaping due to a combination of linear and nonlinear effects, including diffraction, plasma defocusing, and nonlinear self- and cross-phase modulation.
	
	 The variations in both the magnitude and direction of polarization rotation result from a dynamic interplay between the linear and nonlinear phase differences of the two constituent modes of the FPBs, ultimately determining the polarization orientation as described by Eq. (\ref{eq7}). At short propagation distances, the polarization rotation is primarily governed by the difference in the accumulated nonlinear phases of the two components, as the Gouy phases of both modes are negligible and plasma effects are absent. During this stage, the higher intensity of the LCP Gaussian component drives a clockwise rotation of the polarization states in the intense regions of the Poincaré beams. As shown in Fig. \ref{fig6} and Supplementary Video 1 \cite{SuppMat2024}, as the propagation distance increases both modes undergo nonlinear focusing, leading to reshaping of the Gaussian component.  This reshaping induces a buildup of the Gouy phase difference between the two modes, resulting in a counterclockwise rotation of the polarization states. Around $z=85$ cm, the rate of polarization rotation decreases as a balance emerges between the accumulated linear and nonlinear phase contributions of the two components. Beyond this point, the rotation direction reverses due to the higher intensity of the Gaussian component at a radius of approximately 150 $\mu$m, continuing until $z=110$ cm. After $z=110$ cm, the most intense part of the Gaussian mode remains near the propagation axis while the LG mode diffracts, causing the Gouy phase difference to dominate again and leading to a resumption of counterclockwise polarization rotation. The slight differences in the polarization rotation patterns of the star and lemon FPBs can be attributed to subtle variations in the reshaping of their LCP Gaussian components, which arise from the opposite vorticities of the LG modes in the two beams (see Supplementary Video 1 \cite{SuppMat2024}).
	 
	 Additionally, the local amplitude ratio of these modes governs the ellipticity of the polarization states across the beam profile. For example, nearly equal amplitudes of the two modes (e.g., near the nonlinear focus at a radius of approximately 150 µm) result in nearly linear polarization states, while significant amplitude differences lead to pronounced deviations from linear polarization.

	\section{conclusion}
	\label{sec4}
	Through numerical simulations, we investigated the filamentation of fs vector beams. The vector beams are in the form of azimuthal, radial, and spiral cylindrical vector beams (CVBs), and also star and lemon Full Poincar\'e beams (FPBs). All vector beams are considered as vector superpositions of two spatial modes with orthogonal circular polarizations. In the case of the CVBs, the polarization components are in the form of LG modes with opposite topological charges, while for the FPBs the components are a left-circularly-polarized (LCP) Gaussian mode and a right-circularly-polarized (RCP) LG mode with the topological charge of $\pm1$ for the star and lemon polarization topologies, respectively. Our results revealed that the filamentation of all three CVBs is identical since they have equal amplitudes of the constituent components, and the phase difference between the components that determine the polarization profiles of the CVBs has no effect on their filamentation. We also compared the filamentation of the CVBs with a circularly polarized LG beam. All beams exhibited well-known characteristics of filamentation in atmospheric pressure air such as intensity clamping beyond 10$^{13}$ W/cm$^{2}$, and electron densities of more than 10$^{16}$ cm$^{-3}$ over the light filaments. Interestingly, we demonstrated that due to the simultaneous action of both the self- and cross-Kerr effects during the nonlinear propagation of the CVBs, they are more prone to filamentation than the circularly polarized LG beam where only the self-Kerr effect is present.
	 
	We also investigated the filamentation of star and lemon FPBs, and the mutual interaction of their LCP Gaussian and RCP LG modes during the nonlinear propagation and their contribution to the filamentation of the whole beams. Our results indicate that the LCP Gaussian component of the FPBs has the main contribution to the filamentation of these beams since their LG modes are more resilient to the formation of light filaments. We compared the filamentation of these FPBs with an equivalent linearly polarized (LP) Gaussian beam and demonstrated that the LP Gaussian beam more easily forms filament starting from a shorter propagation distance, with a longer length, and a smaller transverse size in comparison to FPBs. 
	
	Furthermore, we studied the evolution of the polarization profile of the vector beams during their filamentation. We demonstrated that the polarization profiles of the CVBs remain substantially unchanged, while the polarization morphologies of the FPBs experience remarkable nonuniform changes because of the difference between the accumulated nonlinear phases of the two components, which is nonuniform due to nonlinear mode reshaping, and the difference between the Gouy phases of the two modes. This is in contrast to the evolution of the profile during the linear propagation where a uniform rotation of the polarization pattern is expected to occur solely due to the variation of the Gouy phase difference between the two polarization components.	

Our findings may have the potential to contribute to advancements in the field of ultrashort pulse filamentation and its applications. These results provide new insights into the application of intense ultrashort pulses, especially in situations where the polarization state of the intense light filaments is crucial such as in THz generation, high-harmonic generation, molecular alignment with ultrashort pulses, and laser-induced material structuring, where the spatially varying polarization profiles of the intense light filaments can be utilized to form sculptured transient currents, achieve non-uniform or cylindrically symmetric molecular alignment patterns, induce engineered material modification patterns, etc. Additionally, we propose that the polarization shaping of the initial femtosecond beam could enable precise control over light filament characteristics like peak intensity, plasma density, and diameter, which are of key importance for numerous applications of femtosecond filamentation.

	\bibliography{references}
		
	\end{document}